\documentclass[twocolumn,prd,showpacs,superscriptaddress,groupedaddress,preprintnumbers,nofootinbib,color]{revtex4-1}
\usepackage{graphicx,amsfonts,amsmath,amssymb,epsfig,color, mathrsfs,latexsym,url,wasysym,float,xfrac}
\newcommand{\be}{\begin{equation}}
\newcommand{\ee}{\end{equation}}
\newcommand{\nn}{\nonumber}
\newcommand{\ba}{\begin{eqnarray}}
\newcommand{\ea}{\end{eqnarray}}
\newcommand{\mpl}{m_{\rm Pl}}

\begin{document}

\title{Higgs in Nilpotent Supergravity: Vacuum Energy and Festina Lente 
}
\author{Amineh Mohseni$^a$}
\author{Mahdi Torabian$^{a,b*}$}
\affiliation{\centerline{$^a$Department of Physics, Sharif University of Technology, Azadi Ave, P.Code 1458889694, Tehran, Iran\\}}
\affiliation{$^{b}$International Centre for Theoretical Physics, Strada Costiera 11, 34151, Trieste, Italy} 

\begin{abstract}
In this note we study supergravity models with constrained superfields. We construct a supergravity framework in which all (super)symmetry breaking dynamics happen in vacuum with naturally (or otherwise asymptotically) vanishing energy. Supersymmetry is generically broken in multiple sectors each of them is parametrized by a nilpotent goldstino superfield. Dynamical fields (the Higgs, inflaton, etc) below the supersymmetry breaking scale are constrained superfields of various types. In this framework, there is a dominant supersymmetry breaking sector which uplifts the potential to zero value. Other sources of supersymmetry breaking have (asymptotically) vanishing contribution to vacuum energy such that supersymmetry is locally restored. Demanding vanishing vacuum energy constrains the structure of the superpotential and Kahler potential; there is a superpotential term for each secluded sector directly interacting with a nilpotent superfield and the Kahler potential must have a shift symmetry along Higgs field directions. This structure is inspired by elements that appear in string theory. We also study the Higgs dynamics during inflation and show that the swampland Festina Lente bound could be realized in this framework.
\end{abstract}

\preprint{SUT/Physics-nnn}
\maketitle

\subsection*{I. Introduction}
Explaining the present state of the universe happens to be laborious from theoretical perspectives. Experiments indicate that we live in a vacuum in which the electroweak symmetry is broken through the Higgs mechanism and the vacuum energy is positive so we live in a de Sitter spacetime \cite{ATLAS:2012yve,CMS:2012qbp,Planck:2018vyg}. The shape of the scalar potential along the Higgs field direction around the local minimum is known: the minimum is displaced from the origin of the field space by around a hundred GeV, parametrically has the same amount of curvature and, if it is non-zero and contributes to the dark energy, it is not elevated by more than around meV$^4$. These mass scales, the Higgs mass parameter and the cosmological constant, are respectively 16 and 120 orders of magnitudes smaller than the Planck units. As input Wilsonian parameters in an effective field theory (without gravity) they suffer from fine-tuning of initial conditions {\it a.k.a} the naturalness problems. Qualitatively, similar ideas and understandings apply to any other scalar in the field space. Extensive attempts were made to explain the fine-tuned electroweak scale and cosmological constant in the string landscape with dense enough number of metastable vacua \cite{Bousso:2000xa,Douglas:2006es,Susskind:2003kw}. 

The string swampland program offers a new perspective to understanding the naturalness issues; not all effective field theories which look consistent from the IR criteria can be UV completed in a theory of quantum gravity \cite{Vafa:2005ui,Ooguri:2006in}. The set of consistent models are basically of measure zero (points) in the vast theory space that are very constrained. Constraints are imposed through numerous conjectures with different levels of rigorousness (see \cite{Brennan:2017rbf,Palti:2019pca,vanBeest:2021lhn}  for review). 
There are swampland conjectures that are directly relevant to the present state of the universe in a de Sitter vacuum (and presumably at some point in the early universe during cosmic inflation). The swampland de Sitter conjecture (SdSC) implies that there cannot exist a meta-stable de Sitter vacuum in asymptotic regions of moduli space \cite{Ooguri:2018wrx,Obied:2018sgi,Agrawal:2018own,Garg:2018reu,Danielsson:2018ztv,Andriot:2018wzk} although short-lived ones in the bulk are allowed through the trans-Planckian censorship conjecture (TCC) \cite{Bedroya:2019snp}. 
Another relevant conjecture is the Festina Lente (FL) bound which in particular puts a lower bound on the mass  of charged particles in de Sitter spacetime \cite{Montero:2019ekk,Montero:2021otb}. These conjectures essentially constrain the shape of the scalar potential. The SdSC and the TCC constrain the slope of the potential and the lifetime of the local minimum. 
The FL bound implies that the Higgs field value is related to the value of vacuum energy. 

An important task in the swampland program in a top-down approach would be to find the largest set or web of swampland conjectures. It helps to construct consistent phenomenological theories that respect those conjectures.  Eventually, a very limited set of theories with particular boundary conditions in the UV survive the swamp as physical theories in the IR. The low energy dynamics of superstrings is described by supergravities. In a bottom-up approach, one might study supergravities which respect swampland conjectures and thus may be UV completed in a quantum gravity theory (see \cite{Katz:2020ewz,Ferrara:2019tmu,Rasulian:2021wny} for recent studies).  The de Sitter-related swampland conjectures constrain the shape of the scalar potential of a supergravity and the location of its minima. 

On the other hand, the vacuum energy in a supergravity is not arbitrary and is bounded  $1\geq \Lambda \geq -3m_{3/2}^2$ (in Planck unit) where $m_{3/2}$ is the gravitino mass. 
Any (super)symmetry breaking effect through non-vanishing $F$ or $D$ term uplifts the potential. The present vacuum favors a tiny a cosmological constant  $\Lambda \lesssim 10^{-120}$. Thus, the scalar potential of a supergravity is constrained; supersymmetry must be broken and the minimum must be lifted. Generically, supersymmetry can be independently broken in multiple secluded sectors. Moreover in each sector, there might be a Higgs-like symmetry breaking dynamics. Unrelated sources of (super)symmetry breaking induce different contributions to the vacuum energy and all must sum up so it is consistent with zero. Besides, the LHC results show no hint of TeV scale  supersymmetric spectrum and implies that supersymmetry, as an indispensable part of string theory, is not part of the low energy physics. Therefore, the dominant source of supersymmetry breaking could be as high as the string scale, {\it i.e.}, $m_{\rm EW} \ll m_{\rm SUSY}\lesssim m_{\rm string}$. 

At lower energies, supersymmetry transformation is  non-linearly realized among component fields \cite{Volkov:1973ix}. However, it can be realized in a linear way through applying constraints on superfields to project out some of the components from the low energy spectrum and allow a subset to propagate \cite{Komargodski:2009rz,Ferrara:2015tyn,Kahn:2015mla,Kallosh:2014via,DallAgata:2015zxp,DallAgata:2016syy,Cribiori:2016hdz,Cribiori:2017ngp}.  

In this paper, we study supergravities with nilpotent superfields. In this framework 
it is possible to have a supersymmetry breaking vacuum with zero or asymptotically zero vacuum energy. These supergravities are known to be the low energy limit of IIB theory with anti-D3 branes where constrained superfields live on their world-volume \cite{Ferrara:2014kva,Kallosh:2014wsa,Bergshoeff:2015jxa,Bergshoeff:2015tra}. Naturally vanishing  energy in supersymmetry breaking vacuum was studied in no-scale models with a particular Kahler structure and superpotential terms \cite{Cremmer:1983bf,Lahanas:1986uc}. Nilpotent supergravities offer an alternative (a closely connected though \cite{DallAgata:2014qsj}) framework with symmetry breaking in vacuum with naturally vanishing energy. We study the Higgs and inflationary dynamics in nilpotent supergravity with constrained superfields. Moreover in this framework, we study the Higgs behavior during inflation especially the expectation value and its relation to the vacuum energy. This model has not been directly obtained from a low energy limit of a string compactification, but it is inspired by elements that appear in string theory. Moreover, we find that the FL bound which is a swampland criterion for quantum gravity can be satisfied in this framework. Demanding vanishing vacuum energy constrains the structure of the superpotential such that it enjoys a scale symmetry (closely related to the no-scale structure as mentioned above). 
Furthermore, the Kahler potential must have a shift symmetry along Higgs field directions.

The structure of the paper is as follows; in section 2 we first briefly review the nilpotent supergravity and present a framework in which vacuum energy vanishes after an inflationary and Higgs dynamics. In section 3, applying this framework, we build a model of Higgs during primordial inflation and study the FL bound. Finally, we conclude in section 4.

\subsection*{II. Nilpotent Supergravity and Vacuum Energy}
In this section, we first briefly review various constrained superfields and show which components are projected out from the low energy spectrum and which are allowed to propagate. Then, we present a framework in which the supersymmetry breaking, a Higgs mechanism and an inflationary dynamics end up in vacuum with vanishing energy. 

\subsubsection*{Review of constrained superfields}
We start with a chiral superfield $X$ which is defined through the constraint $\bar{\cal D}_{\dot\alpha} X=0$. It has an expansion in Grassmann coordinates
\be X = \phi_{\rm x} + \sqrt2\theta \psi_{\rm x} + \theta^2 F_{\rm x},\ee
where component fields are functions of $x_\mu+i\theta\sigma_\mu\bar\theta$. The non-zero auxiliary component $F_{\rm x}$ signals supersymmetry breaking, the mass spectrum gets split and the fermionic component is (fully or partially in the presence of other sources of supersymmetry breaking) the goldstino which is absorbed by gravitino. Supersymmetry is then realized through non-linear transformations among component fields. It can be realized linearly though, if the superfield is constrained even further as $X^2=0$, namely, it is nilpotent of degree 2. It implies that the components are no longer independent through 3 conditions on components $\phi_{\rm x}^2=\phi_{\rm x}\psi_{\rm x}=0$ and $\phi_{\rm x} = \psi_{\rm x}^2/2F_{\rm x}$. The complex scalar component is completely removed from the spectrum below the supersymmetry breaking scale, and in the unitary gauge (gauge fixing of local fermionic symmetry through $\psi_{\rm x}=0$) we find $\phi_{\rm x}=0$. We call $X$ the goldstino supermultiplet. 

The goldstino superfield $X$ can be used to constrain other superfields and remove components from the spectrum. Given another chiral superfield $\Phi$, the constraint $X\Phi=0$ removes the scalar component of $\Phi$. The fermionic component is projected out through the constraint $X\bar X{\cal D}_\alpha\Phi=0$. If we demand that $\bar{\cal D}_\alpha(X\bar\Phi)=0$, then the fermionic and the auxiliary components will be removed. Furthermore, if we impose the constraint $X(\Phi\pm\bar\Phi)=0$, then either of scalar or pseudoscalar part of the complex scalar component of real fields $(\Phi+\bar\Phi)$ or $(\Phi-\bar\Phi)/2i$ will be removed besides the fermionic and auxiliary components. We note that the latter two constrained superfields have no $F$-term contribution to the scalar potential.
For detailed and extensive discussions see the original papers \cite{Komargodski:2009rz,Ferrara:2015tyn,Kahn:2015mla,Kallosh:2014via,DallAgata:2015zxp,DallAgata:2016syy,Cribiori:2016hdz,Cribiori:2017ngp}. Thus, below the supersymmetry breaking scale, we write an effective supergravity theory of constrained superfields such that only some of components contribute to low energy dynamics. 

\subsubsection*{Supersymmetry breaking and vanishing vacuum energy}
The $F$-term scalar potential of a supergravity including chiral superfields is
\be V_F = e^K(K^{I\bar J} D_IW D_{\bar J}\overline W-3|W|^2),\ee
where the Kahler covariant derivative is $D_IW=(\partial_I+\partial_IK)W=F_I$, and $K^{I\bar J}$ is the inverse of the Kahler metric $K_{I\bar J}=\partial_I\partial_{\bar J}K$ (such that $K^{I\bar K}K_{\bar KJ}=\delta^I_J$). Supersymmetry is broken at some high scale, below which have an effective supergravity theory including a goldstino multiplet $X_0$. Given the following structure for the superpotential and Kahler potential \cite{DallAgata:2014qsj}, the energy in the supersymmetry breaking vacuum is vanishing 
\ba \label{structure} W &=& (1+\sqrt 3 X_0)W_0,\cr
 K &=& X_0X_0^*. \ea
The scale of supersymmetry breaking is fixed in the UV (through moduli stabilization effects, flux, etc) by a constant $W_0$
\be |\partial_{0}W|^2 = 3|W_0|^2 = 3m_{3/2}^2.\ee

This structure is closely related to the well known no-scale structure \cite{DallAgata:2014qsj}. Given the no-scale Kahler potential and superpotential 
\ba&& K=-3\, \ln\left(T+\bar{T}\right),\\
&& W=W_0=\text{const}, \ea
with change of variables as $Z\equiv\sfrac{(2T-1)}{(2T+1)}$ and a Kahler transformation we find
\ba &&K=-3\,\ln\left(1-|Z|^2\right),\\
&&W=W_0(1-Z)^3. \ea
If we assume that $Z$ is a nilpotent superfield, $Z^2=0$, we can expand the above equations around $Z=0$. Upon a redefinition, $X_0\equiv -\sqrt{3}Z$,  we arrive at \eqref{structure}. Therefore, the above structure enjoys a scale symmetry which typically appears in string compactifications \cite{ferrara1994mass}. This scale symmetry determines coefficients of the superfield $X_0$ in the linear structure of the superpotential. Scale symmetry is not in general respected by quantum corrections and question of  stability of this structure against quantum corrections is a subject for further study.

The above framework can be generalized to include other superfields, say $\Phi$, in a generic superpotential
\be W \supset (1+\sqrt 3 X_0)[W_0+ w(\Phi)][1 + {\cal O}(\Phi)],\ee
 where $w$ is any singlet composite operator, and $\cal O$ stands for higher order singlet induced by supergravity effects respectively of with an $R$-charge +2 and 0. In order not to contribute to vacuum energy, $w$ and ${\cal O}$ must be either functional of constrained superfields with projected out $F$ component or superfields with vanishing $F$ component in the vacuum.
A generic Kahler potential in supergravity includes higher-order Planck-suppressed terms. For instance, unless $X_0 {\cal K}(\Phi,\bar\Phi)=0$, we can consider 
\be K \supset (X_0+X_0^*){\cal K}(\Phi,\bar\Phi),\ee
where ${\cal K}$ is a composite real superfield. In order to suppress its contribution to the vacuum energy, $\cal K$ must be vanishing in the vacuum $\langle {\cal K}\rangle=0$.

Furthermore, the above framework can be generalized to include multiple nilpotent superfields. Supersymmetry can be broken in $N$ sectors which are is parametrized by multiple goldstino superfields $X_I$, and order parameters $f_I$. Then, the superpotential includes the following contribution
\ba W \supset {\sum_{I=1}^N}X^I\big[f^I+{\cal F}^I(\Phi)\big],\ea
where $\cal F$ is a holomorphic function of constrained superfields $\Phi$. 
The Kahler potential $K = K(\Phi,\bar\Phi)$ is an arbitrary function of $\Phi\bar\Phi$ or $\Phi\pm\bar\Phi$. Superfields are constrained so that they have vanishing $F$-terms. Only goldstino superfields have non-vanishing $F$-terms and contribute to the vacuum energy. 
Then, the positive semi-definite $F$-term contribution to scalar potential is 
\ba V&\supset&  e^{K}  \textstyle\sum_I K^{I \bar I}\big|f^I+{\cal F}^I\big|^2. \ea
In this work, we are interested in models in which the above contributions to vacuum asymptotically approach zero, namely, SUSY is restored in all except one sector; the dominant and non-vanishing source of SUSY breaking is from the $X_0$ sector.

\subsubsection*{The Higgs mechanism and vanishing vacuum energy}
In this part, we study phase transition via Higgs dynamics in the framework of constrained supergravity with nilpotent superfields which ends with vanishing vacuum energy. 
Although the results of this section are general and hold for any Higgs dynamics, we concentrate on the Standard Model Higgs mechanism. We consider two Higgs doublets $H_u$ and $H_d$ which are constrained as $\bar{\cal D}_{\dot\alpha}(X\bar H_u)=\bar{\cal D}_{\dot\alpha}(X\bar H_d)=0$, so that the Higgsinos are heavy and decoupled from the spectrum and the Higgs $F$-terms are vanishing in the unitary gauge (see \cite{Li:2020rzo} for an earlier attempt. In this work we include all supergravity terms in a more general framework). 
	
We build a model with two goldstino superfields $X_0$ and $X_1$. The superpotential is 
\ba W &=& (1+\sqrt 3 X_0)W_0\big[1+{\cal F}(H_u\!\cdot\! H_d)\big] \cr &+& X_1\big[f_1+{\cal F}_1(H_u\!\cdot\! H_d)\big], \ea
where $\cal F$ and ${\cal F}_1$ are polynomials of singlet $H_u\cdot H_d$. 
An unsuppressed holomorphic mass term for the doublets (the leading order $\mu$ term) can be forbidden by an $R$ symmetry such that $R(H_uH_d)=0$, or a (geometric) discrete symmetry. It can also be forbidden by scale invariance in a UV theory as in Casas-Munoz model in string theory. The same symmetry forbids interactions between Higgses and $X_0$ at the leading order. It is, however, induced through higher dimensional operators. When geometric moduli are stabilized, any $R$-symmetry or geometric symmetry related to the compact manifold is broken. With a non-zero $W_0$, the higher-order supergravity effects can generate an effective $\mu$ parameter through a dimension 5 operator. 

Moreover, we consider a shift symmetric Kahler potential along the Higgs direction (see \cite{Hebecker:2012qp,Hebecker:2013lha} for realizations in string theory)
\ba &&H_u \rightarrow H_u + {\cal C},\cr
&& H_d \rightarrow H_d -\bar{\cal C} .\label{n-shiftsymm}\ea
As we will see, the shift symmetry helps to control the higher-order supergravity terms in the Kahler potential.
Thus, the Kahler potential is
\ba K\! &=&\! |X_0|^2\! +\! |X_1|^2\!+\! |H_u\!+\!\bar H_d|^2\! +\! (X_0\!+\!X_0^*){\cal K}(|H_u\!+\!\bar H_d|^2),\nn\\ \ea
where we included the higher order terms in $\cal K$ which is a function of the singlet $|H_u\!+\!\bar H_d|^2$.
	
The scalar potential, induced by $F$ and $D$ terms, is 
\ba	V&=& e^{ |H_u + \bar H_d |^2}\Big[\left|  f_{1} + {\cal F}_{1}\right|^2\cr
&+& \!\left|1+{\cal F}\right|^2\!W_0^2\big((1-2|H_u + \bar H_d |^2(\partial{\cal K})^2)^{-1}\big(\sqrt{3}+{\cal K}\big)^2\!-\!3\big)\Big]\cr
&+& \sfrac{1}{8}(g^2+g'^2)(|H_u|^2-|H_d|^2)^2,\ea
where $\partial{\cal K}$ is derivative with respect to the argument of function ${\cal K}$. We note that the scalar potential is positive semi-definite. Moreover, we note that $V_{X_0}-3|W|^2\sim {\cal K}$.

We expand the Higgs doublets and use the isospin gauge symmetry to set $H^{+}_u=0$ and consequently $H^-_d=0$. The phase of $f_1$ is absorbed into the relative phase of the Higgses. Then, thorough a hypercharge rotation we completely remove the phases so that the neutral components are real and positive. Thus, we can expand the functions of Higgs doublets as
\ba {\cal F} &=& -cH_u^0H_d^0 +\cdots,\cr
{\cal F}_1 &=& -c_1H_u^0H_d^0 +\cdots,\cr
{\cal K} &=& k(H_u^0 - H_d^0)^2+\cdots,\ea
where dots stand for higher order Planck suppressed terms, and $c,c_1,k$ are order one constants. The scalar potential is
	\ba V&=& e^{(H_u^0-H_d^0)^2}\big[(f_1-c_1 H_u^0H_d^0 +\cdots)^2\cr
      &&\qquad\qquad\ +  W_0^2 (1-c H_u^0 H_d^0+\cdots)^2\times\cr &&\qquad\qquad\qquad\quad\times k(H_u^0-H_d^0)^2 \big(2(3k+\sqrt3)+\cdots\big)\big]\cr      &+& \sfrac{1}{8}(g^2+g'^2)(H_u^{0\,2}-H_d^{0\,2})^2.
 \ea
To the leading order, neglecting the Planck suppressed modifications, the minimum of the potential is
\be H_u^0=H_d^0 = v \approx \sqrt{\frac{f_1}{c_1}}.\ee
Higher order terms, displace the location of the minimum of order $v^8$ in Planck units. 
Apparently, the scalar potential vanishes at this minimum and this model gives electroweak symmetry breaking in a vacuum with zero energy. The supersymmetry breaking vacuum is lifted to the zero energy value via $X_0$ part and the vanishing contribution from Higgs dynamics is guaranteed through the $X_1$ coupling in the superpotential and the shift symmetry in the Kahler potential. The above analysis could be applied to any other Higgs-like dynamics in other sectors. We not that it is mandatory to have Higgs multiplets in vector-like representation of gauge group. In a supersymmetric setup, holomorphic structure of the superpotential and chiral anomaly constrain need   vector-like Higgses. We find that must be also so in a supergravity framework with vanishing vacuum energy.

\subsubsection*{An inflationary dynamics ending in a Minkowski vacuum}
	In this framework of constrained supergravity, a minimal model for inflationary dynamics can be constructed by an inflaton superfield $\phi$ that satisfies $X(\phi+\phi^*)=0$ \cite{Ferrara:2015tyn,Carrasco:2015iij}. This constraint keeps only a scalar field dynamics. This in particular could be useful in models with an exact or approximate shift symmetry. The general forms of a (minimal) superpotential and Kahler potential are 
		\ba W &=& (1+\sqrt 3 X_0)[W_0+{\cal G}(\phi)] + X_2(f_2+\!A_2e^{-a_2\phi}),\quad \ \\
	K &=& |X_0|^2 + |X_2|^2+ \sfrac{1}{2} (\phi+\phi^*)^2,\ea
where $\cal G$ is an arbitrary function of $\phi$. The Kahler potential is shift symmetric in the inflaton direction. However, this symmetry is broken by non-perturbative terms in the superpotential. It is minimal in the sense that there is a single supersymmetry breaking/uplifting sector besides  the inflationary sector.
Given that $  |D_{X_0}W|^2  -3|W|^2$ is vanishing, the scalar potential is
	\be V = |f_2+A_2e^{-a_2\phi}|^2 = A_2^2+f_2^2+2A_2 f_2 \cos(a_2 \varphi),\ee 
where $\varphi$ is the imaginary part of $\phi$. We note that in this framework there is no additional scalar field (the supersymmetric partner of inflaton {\it a.k.a.} siflaton/saxion) that needs stabilization.  

At the end of inflation, the scalar potential is asymptotically approaching zero if
\be A_2=-f_2.\ee
The scalar potential will be 
	\be 
		V=V_{\rm inf} \left[1-\cos\left(a_2 \varphi\right)\right],
	\ee
where $V_{\rm inf}=2f_2^2=2A_2^2$ sets the scale of natural axionic inflation. Alternative inflationary potentials could be constructed in the framework of nilpotent supergravity \cite{Kallosh:2014via,Ferrara:2014kva,DallAgata:2014qsj}. At the end of inflation, we end up in a Minkowski vacuum with high-scale supersymmetry breaking fixed by $W_0$.

\subsection*{III. Higgs during Inflation: the FL bound}
In this section, we study the Higgs dynamics at the time of cosmic inflation. In particular, we are interested in building a model in which the swampland FL bound is respected during cosmic evolution with positive vacuum energy. The FL bound is basically a lower bound on the mass of a particle of charge $q$ in a de Sitter spacetime with energy density $V$
\be m^2\geq (gq)V^{1/2},\ee
where $g$ is the $U(1)$ coupling constant. In the present universe the bound  in the visible sector is satisfied by a huge margin. However, the bound could have non-trivial implications for the Higgs dynamics in any earlier stage of cosmic inflation with large vacuum energy. As the Higgs field value fixes the mass of charged particles, it must have larger values during a primordial inflation. It has also implications for the hidden sector at the early and late universe (see \cite{Ban:2022jgm,Lee:2021cor,Guidetti:2022xct,Montero:2022jrc} for phenomenological studies).

In a minimal model, we consider three goldstino nilpotent superfields, $X_0, X_1, X_2$, two constrained Higgs doublet superfields and a constrained inflaton superfield as in the previous section. Considering all supergravity terms allowed by symmetries and a Higgs shift symmetry, the superpotential and the Kahler potential are
\ba W &=& (1+\sqrt 3 X_0)\big[W_0+{\cal G}(\phi)\big]\big[1+{\cal F} (H_u\!\cdot\! H_d)\big]\cr  
&+& X_1[f_1+A_1e^{-a_1\phi}+{\cal F}_1(H_u\!\cdot\!H_d)] \cr &+&  X_2(f_2+\!A_2e^{-a_2\phi}), \\
K &=& |X_0|^2 + |X_1|^2+ |X_2|^2+ |H_u\!+\!\bar H_d|^2 + \sfrac{1}{2} (\phi+\phi^*)^2 \cr
&+&  (X_0\!+\!X_0^*){\cal K}(|H_u\!+\!\bar H_d|^2)+ (\phi\!+\!\phi^*){\cal J}(|H_u\!+\!\bar H_d|^2).\nn\\\ea
Here for the sake of concreteness, we consider an approximate shift symmetry in the inflaton sector. Alternative models of inflation can be considered without changing the main conclusion. Functions ${\cal F, F}_1,\cal J, K$ are arbitrary functions of Higgs doublets that can be expanded as 
\ba {\cal F} &=& cH_u \cdot H_d +\cdots,\cr
	{\cal F}_1 &=& c_1H_u \cdot H_d +\cdots,\cr
	{\cal K} &=& k |H_u + \bar H_d|^2+\cdots,\cr
	{\cal J} &=& j |H_u + \bar H_d|^2+\cdots,\ea
	where dots stand for Planck suppressed supergravity terms, and $c,c_1,k,j$ are order one constants. 
Moreover, $\cal G$ is an arbitrary function of $\phi$.  Without loss of generality we take it as ${\cal G} = A_0e^{-a_0\phi}$ that non-perturbatively break the shift symmetry. We postpone answering the question of whether the string inspired scaleless structure of the supepotential is stable against quantum corrections to a future work. We have taken into account all possible higher order operators given this structure.

The scalar potential is read as follows 
\ba
V&=& e^{ |H_u + \bar H_d |^2}\Big[\left|  f_{1} + A_1 e^{-a_1\phi } + c_{1} H_{u}\!\cdot\!H_{d} +\cdots\right|^2\cr
&&\qquad\qquad+\left|f_2+A_2 e^{-a_2\phi}\right|^2 \cr
&&\qquad\qquad+\left|(W_0+A_0 e^{-a_0 \phi})(1+c H_u\!\cdot\! H_d+\cdots)\right|^2\times\cr
&\times&\!\!\!\Big(\frac{(1\!-\!2 j^2 |H_u\!+\!\bar{H}_d|^2\!+\!\cdots)(\sqrt{3}\!+\!k |H_u\! +\! \bar H_d |^2\!+\!\cdots)^2}{1-2 (k^2+j^2) |H_u+\bar{H}_d|^2+\cdots}\!-\!3\Big)\!\Big]\cr
&+& \sfrac{1}{8}(g^2+g'^2)(|H_u|^2-|H_d|^2)^2,\ea
We expand the Higgs doublets and use the $SU(2)_L$ gauge symmetry to set $ H^{+}_u=0=H_d^-$. We show the phase difference between the two Higgses  by $ \theta $, thus
\ba |H_u + \bar H_d |^2 &=& |H^0_u-e^{-i\theta} H^0_d|^2,\\
H_u\cdot H_d &=& - e^{i\theta} H^0_u H^0_d,\ea
where  $H^0_u$ and $H^0_d$ are real and positive. Moreover, we absorb the phase of $ f_{1} + A_1 e^{-i a_1\varphi}$ into the relative phase of the Higgses and take $f_1,A_1$ real. For that, we set the parameter $\theta$ to be the function of the scalar modulus
\be \label{n-theta} \theta\left(\varphi\right)=\tan^{-1}\bigg[\frac{\text{Im}\left( f_{1} + A_1 e^{-i a_1\varphi}\right)}{\text{Re}\left( f_{1} + A_1 e^{-i a_1\varphi}\right)}\bigg], \ee
where $ \varphi $ is the imaginary part of the scalar component of $\phi$. In the minimum of the potential the product $H_u^0H_d^0$ must also be real and positive. We set the Higgs phases to zero through a hypercharge rotation. 

The scalar potential as a function of three scalar fields $V=V(\varphi,H_u^0,H_d^0)$ is
	\ba \label{scalarpot} V&=&e^{|H^0_u-e^{-i\theta} H^0_d|^2}\Big[\big(G_1(\varphi)- c_{1} H^0_{u}H^0_{d}\!+\!\cdots\big)^2 + G_2(\varphi)^2 \cr 
	&&\qquad\qquad\qquad+G_0(\varphi)^2\big|1-c e^{i \theta} H^0_u H^0_d+\cdots\big|^2\times\cr
	&&\times\Big(K^{-1}(H^{0}_u,H^{0}_d)(\sqrt{3}\!+\!k |H^0_u-e^{-i\theta} H^0_d|^2\!+\!\cdots)^2-3\Big)\Big]\cr
	&+&\sfrac{1}{8}(g^2+g'^2)(H_u^{0\,2}-H_d^{0\,2})^2,\ea
where we have defined
\ba G_0(\varphi)^2 &=& W_0^2+A_0^2+2W_0A_0\cos(a_0 \varphi),\\
G_1(\varphi)^2 &=& f_{1}^2 + A_1^2 + 2 f_{1} A_1  \cos(a_1 \varphi),\\
G_2(\varphi)^2 &=& f_{2}^2 + A_2^2 + 2 f_{2} A_2  \cos(a_2 \varphi),\\
K^{-1}(H^{0}_u,H^{0}_d) &=& \frac{1-2 j^2|H^0_u-e^{-i\theta} H^0_d|^2+\cdots}{1\!-\!2 (k^2\!+\!j^2) |H^0_u\!-e^{-i\theta} H^0_d|^2\!+\cdots}.\quad\ \ea
Clearly, given $k>0$ the potential is positive semi-definite. 
The minimum of the potential is found by 
\begin{equation}\label{n-higgsopt}
	\frac{\partial V}{\partial H_u^0}=\frac{\partial V}{\partial H_d^0}=0.
\end{equation}
If the following condition among parameters is imposed
\be\label{n-conditions} G_2(\varphi)\ll G_1(\varphi)<c_1,\ee
a  non-trivial electroweak symmetric breaking solution is found as
\ba\label{n-H2} H_u^{0\,2}=H_d^{0\,2}&=&c_1^{-1} G_1(\varphi)-c_1^{-2}(1\!-\!\cos\theta)G_2^2(\varphi) \cr
\!&+&\!{\cal O}\sfrac{(m_{\rm SUSY}^6}{\mpl^6}),\qquad \ea
where we kept terms up to ${\cal O}(G_2^2)$ and Planck-suppressed terms starting with ${\cal O}(m_{\rm SUSY}^6/\mpl^6)$ appear due to quantum corrections. These terms can be neglected during inflation and exactly vanish at the minimum as we argue in the following. We note that (see eq \eqref{scalarpot}) due to the shift symmetry of the Kahler potential (see eq \eqref{n-shiftsymm}) the contribution to the vacuum energy from higher order terms proportional to $k$ exactly vanishes along minimum of the potential $\varphi=\theta=0,\ H_u^0=H_d^0=v$. We remind that for $k>0$ the potential is positive semi-definite. These higher order terms do not displace the minimum and has vanishing contribution to the vacuum energy.

Along this direction in field space, the Higgses contribution to vacuum energy is ${\cal O}(G_2^4)$ and the scalar potential as a function of $\varphi$ up to ${\cal O}(G_2^2)$  is
\ba\label{n-Vsphi}
	V\left(\varphi\right)&=&e^{2c_1^{-1}(1-\cos\theta)G_1(\varphi)}G_2^2(\varphi) \approx G_2^2(\varphi)\cr
	&=&f_{2}^2 + A_2^2 +2 f_{2} A_2  \cos( a_2 \varphi).
\ea
The scalar potential approaches zero at the end of inflation, {\it i.e.} $G_2(0)=0$, provided that
\be A_2=-f_2.\ee
Then, the potential drives a natural inflation along the axion direction  
\begin{equation}\label{n-infpot}
	V\left(\varphi\right)=V_{\rm inf} \left[1-\cos(a_2 \varphi)\right],
\end{equation}
where $V_{\rm inf} = 2f_2^2=2A_2^2$.
We parametrize the electroweak vacuum as $ \langle H^{0}_u\rangle \equiv v_u/\sqrt{2}, \langle H^{0}_d\rangle \equiv v_d/\sqrt{2}$ and $v_u^2+v_d^2=v_{\rm EW}^2$. The present universe is in the electroweak broken vacuum $v=v_{\rm EW}$ at $\varphi =0$ and given that $G_2(0)=0$ equation \eqref{n-H2} implies
\begin{equation}\label{n-ewBCresult}
	G_1(0)=\frac{c_1}{4} v_{EW}^2,
\end{equation}
which consequently means 
\begin{equation}\label{n-Af1relnnn}
	A_1 + f_1 = \frac{c_1}{4} v_{EW}^2.
\end{equation}
It can be used to fix $G_1(\varphi)$ as
 \ba\label{G1} G_1(\varphi)\! &=&\! \big[f_1^2\!+\!(f_1\!-\!\frac{c_1 }{4}v_{EW}^2)(f_1\!-\!\frac{c_1 }{4}v_{EW}^2\!-\!2 f_1 \cos(a_1 \varphi))\big]^{1/2}\cr
 &=& \big[2 f_{1}^2 (1-\cos(a_1 \varphi))\big]^{1/2}+\mathcal{O}\left(\sfrac{v_{EW}^2}{f_1^2}\right).\ea
The Higgs vacuum expectation value is found as
\ba v^2\!&=&\!\frac{4}{ c_{1}}\big[f_1^2\!+\!(f_1\!-\!\frac{c_1 }{4}v_{EW}^2)(f_1\!-\!\frac{c_1 }{4}v_{EW}^2\!-\!2 f_1 \cos(a_1 \varphi))\big]^{1/2}\cr
&-&\frac{4}{c_1^{2}}(1-\cos\theta)G_2^2(\varphi)\cr
&=& \frac{4}{ c_{1}} \big[2 f_{1}^2 (1-\cos(a_1 \varphi) ) \big]^{1/2}-\frac{4}{c_1^{2}}(1-\cos\theta)G_2(\varphi)^2\cr &+&\mathcal{O}(\sfrac{v_{EW}^2}{f_1^2})\label{n-higgsvevfinal},\ea 
which is clearly axion (inflaton) dependent. 

The FL bound for the electron, the lightest charged particle in the visible sector, implies
\begin{equation}
	v_d^4 \gtrsim 8 \frac{q^2 g^2}{y_e^4} V\left( \varphi \right),
\end{equation}
where $y_e$ is the Yukawa coupling of the electron. 
We use equations \eqref{n-infpot} and \eqref{n-higgsvevfinal} in the above inequality to find
\begin{equation}\label{n-flinequ}
	\frac{f_1^2}{c_1^{\,2}}\left[1-\cos\left(a_1\varphi\right)\right] \gtrsim \frac{g^2 q^2 V_{\rm inf}}{y_e^4}\left[1-\cos\left(a_2 \varphi\right)\right].
\end{equation}
The FL bound will be satisfied  if the Higgs vacuum expectation value monotonically increases with $ \varphi  $, so that we have a large vacuum expectation value for the Higgs field in the inflationary era. 

Moreover, the condition \eqref{n-conditions} together with \eqref{n-Vsphi}, \eqref{n-infpot} and \eqref{G1}  imply
\begin{equation}\label{n-simpinequ}
	V_{\rm inf} [1-\cos(a_2 \varphi)]\ll 2 f_{1}^2 [1-\cos(a_1 \varphi)]<c_1^2.
\end{equation}
We note that as $y_e\sim 10^{-6}$ the left hand side inequality is satisfied automatically if \eqref{n-flinequ} is satisfied. 

Demanding \eqref{n-flinequ} and \eqref{n-simpinequ} hold in the inflationary regime where $\varphi\sim \frac{\pi}{a_2}$ we find for $f_1>0$ the following constraints
\ba\label{n-final2}
	f_{1}<\frac{ c_{1}}{2}\quad{\rm and}\quad  f_2^2<\frac{y_e^4 f_{1}^2 }{2 q^2 g^2  c_{1}^{2}}\sin^2\Big(\frac{\pi a_1 }{2 a_2}\Big),
\ea
which are satisfied for
\begin{equation}
	|f_2|\lesssim y_e^2  |f_1| \ll |f_1|.
\end{equation}
Therefore, the FL bound can be satisfied in this framework in a broad range in the parameter space. 

Finally, we would like to comment on the implications of the weak gravity conjecture (WGC) and the no global symmetries conjecture (NGSC) for our model. The WGC imposes a constraint on the axion decay constant $f$ through
\be f S \lesssim 1,\ee
where  $S$ is the corresponding instanton action. In this framework, this constraint implies a bound on superpotential coefficients
\be a_0,a_1,a_2 \gtrsim 1. \ee
In fact, it is a bound on the breaking scale of the axionic shift symmetry. This implies a limitation on the duration of inflation as so do other swampland conjectures. In passing we also note the the shift symmetry in the Higgs sector \eqref{n-shiftsymm} is broken at the EW scale through perturbative superpotential and therefore is not in conflict with the NGSC.

\subsection*{IV. Conclusion}
In this paper we presented a framework of supergravity models with constrained superfields of different types in which (super)symmetry breaking naturally end in a vacuum with zero energy. In particular, we considered supersymmetry breaking at an arbitrary scale, electroweak symmetry breaking and inflation. Supersymmetry can be broken in multiple secluded sectors at different scales, each parametrized by a goldstino superfield. All other symmetry breaking/inflationary dynamics happen in each of those sectors. Supersymmetry could be restored in some of them, however, there is a dominant one through which supersymmetry remains unbroken. This is the uplifting sector to the zero value of vacuum energy. 
All  other (super)symmetry breaking sectors 
end up independently giving zero contribution to vacuum energy. It is guaranteed through a particular structure of superpotential and Kahler potential which are motived by string theory constructions. 

We also studied the swampland FL bound for the Standard Model Higgs during inflation. We found that the bound could be easily satisfied in the electroweak sector in the wide region in the parameter space. This study could be generalized to any other Higgs-like symmetry breaking dynamics in other dark sectors. The FL bound also has implications for strong interactions which is postponed to a future work. 
 
 \paragraph*{Acknowledgements} This work is supported by the re- search deputy of Sharif University of Technology. MT is thankful to  the HECAP section of ICTP for hospitality during the final stages of this work.

$^*$ Email: {\tt mahdi.torabian@sharif.ir}

\end{document}